AGU
ADVANCING EARTH AND SPACE SCIENCES

# Geophysical Research Letters®

## JIRAM Observations of Volcanic Flux on Io: Distribution and Comparison to Tidal Heat Flow Models

M. Pettine[1] , S. Imbeah[1,2] , J. Rathbun[1,3] , A. Hayes[1], R. M. C. Lopes[4] , A. Mura[5] , F. Tosi[5] , F. Zambon[5], and S. Bertolino[1]

[1]Cornell University, Ithaca, NY, USA, [2]Malin Space Science Systems, San Diego, CA, USA, [3]Planetary Science Institute, Tucson, AZ, USA, [4]Jet Propulsion Laboratory, California Institute of Technology, Pasadena, CA, USA, [5]INAF-IAPS Istituto Nazionale di Astrofisica-Istituto di Astrofisica e Petrologia Spaziali, Rome, Italy

**Abstract** Juno has allowed clear, high-resolution imaging of Io's polar volcanoes using the Jovian Infrared Auroral Mapper (JIRAM) instrument. We have used data from JIRAM's M-band (4.78 μm) imager from 11 Juno orbits to construct a global map of volcanic flux. This map provides short-term insight into the spatial distribution of volcanoes and the ways in which high- and low-latitude volcanoes differ. Using spherical harmonic analysis, we quantitatively compare our volcanic flux map to the surface heat flow distribution expected from models of Io's tidal heat deposition (summarized in de Kleer, Park, et al. (2019, https://doi.org/10.26206/d4wc-6v82)). Our observations confirm previously detected systems of bright volcanoes at high latitudes. Our study finds that both poles are comparably active and that the observed flux distribution is inconsistent with an asthenospheric heating model, although the south pole is viewed too infrequently to establish reliable trends.

**Plain Language Summary** Our study uses data from an infrared camera on Juno called the Jovian Infrared Auroral Mapper (JIRAM) to image Io, the innermost Galilean moon of Jupiter. Io is the most volcanically active body in the Solar System. Io's volcanoes are powered by both the extreme tides from Jupiter and the gravitational interactions between it and Jupiter's other moons. These tides generate friction inside Io. Simulations of Io's interior suggest that, depending on how deep that friction is being generated, the surface heat flow will be higher in certain areas (de Kleer, Park, et al., 2019, https://doi.org/10.26206/d4wc-6v82). Using JIRAM, we have mapped where volcanoes are producing the most power and compared that to where we expect higher heat flow from the interior models. Our map doesn't agree with any of these models very well. JIRAM observed more volcanic activity at the poles than we expected to see based on previous observations. However, since the south pole was only observed twice, it's possible that these observations don't represent the average volcanic activity of the south pole. Very bright volcanoes that may have been continuously active for decades were also imaged during these Juno fly-bys, some of which are nearer the poles than the equator.

## 1. Introduction

Io, the innermost Galilean moon of Jupiter and the most volcanically active body in the Solar System, is a unique resource for studying volcanism outside Earth and tidally-generated volcanism. Despite its importance in the development of the worlds of the outer Solar System (and beyond), tidal heating remains relatively unconstrained. In fact, the existence of volcanism powered entirely through extreme tidal forces and strong orbital resonances was only proposed and discovered in the last 50 years (Morabito et al., 1979; Peale et al., 1979). Today, many basic questions regarding Io's volcanism remain unanswered: do the types and intensities of volcanic hotspots depend on latitude? And can the spatial distribution of volcanism constrain internal heat dissipation? In this work, we use the Jovian Infrared Auroral Mapper (JIRAM) (Adriani et al., 2017) instrument aboard the Juno spacecraft to address these questions. Juno is in polar orbit around Jupiter, and as such, is the first spacecraft to provide clear, complete, and direct views of Io's poles. Using JIRAM aboard Juno allows for robust analysis of latitudinal variations in Io's volcanism. Several studies have suggested that the dominant eruption styles on Io vary with latitude: longer lived eruptions at low latitudes and shorter duration, higher volume eruptions at higher latitudes (Howell et al., 2001; Lopes et al., 2004; McEwen et al., 2000, 2004). The number of volcanoes and overall volcanic activity have also been reported to be lower at the poles than the equator (Davies et al., 2024; Milazzo et al., 2005; Veeder et al., 2012). While their definition of "persistent" differed from the earlier definitions of "long-lived," de Kleer et al. (2016b) found the persistent hotspots occurred primarily at low latitudes. However, due to the viewing geometry, ground-based observation of Io is systematically biased toward the equator.

19448007, 2024, 17, Downloaded from https://agupubs.onlinelibrary.wiley.com/doi/10.1029/2023GL105782 by Cornell University Library, Wiley Online Library on [23/09/2024]. See the Terms and Conditions (https://onlinelibrary.wiley.com/terms-and-conditions) on Wiley Online Library for rules of use; OA articles are governed by the applicable Creative Commons License

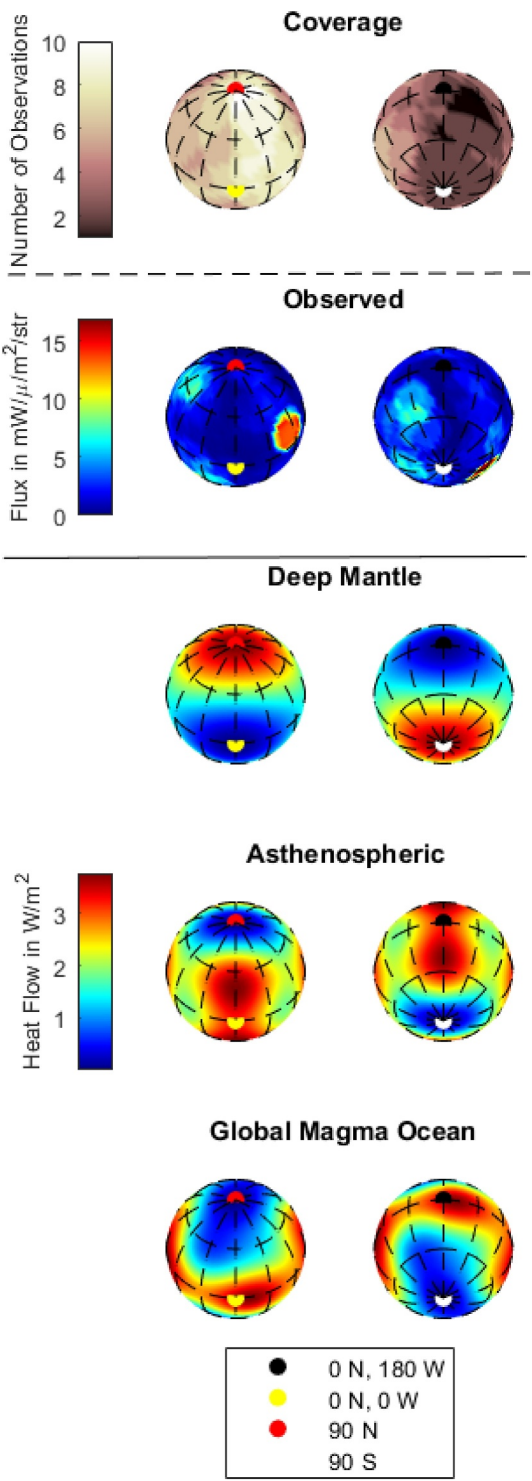

**Figure 1.**

Furthermore, from Earth, Io is only viewable in reflected daylight or in eclipse. Eclipse images are biased toward the sub-Jovian hemisphere and daylit images compete with reflected sunlight. Voyager and Galileo were able to take the first high-resolution images of Io, but they traveled through the same equatorial plane and shared the ground-based bias toward low latitudes; pre-Juno images of Io's poles were sparse and incomplete. This means that Io's polar volcanoes have only been imaged at highly inclined angles, creating a much higher uncertainty in measurements of radiation coming from volcanic hotspots, leading to uncertainties in both intensity and size. This makes Juno especially crucial since it is in polar orbit around Jupiter. While its observations of Io are only taken sporadically due to its trajectory, it nevertheless has, for the first time, imaged the poles directly. We have analyzed 4 years of data from JIRAM, an infrared imager that has both a spectrometer and a camera with L- and M-band filters. We have used its M-band (4.78 μm) observations to compile a global map of volcanic flux on Io (Figure 1).

We point out key interesting features that this map reveals, including confirming bright, high-latitude, long-lived systems of volcanoes that have been previously detected (de Kleer et al., 2014). We compare our map to modeling data for surface heat flow using the most extreme end members of depth (shallow vs. deep mantle heating, from Segatz et al. (1988)) and for the presence of a globally complete subsurface magma ocean (adapted in de Kleer, Park, et al. (2019) using Tyler et al. (2015), Matsuyama et al. (2018)) (Figure 1). We use spherical harmonic analysis to quantitatively study the distribution of volcanic hotspot flux. While our map agrees with previous studies that suggest that low-to-mid-latitude areas see the highest areas of volcanic activity (Davies et al., 2024; de Kleer et al., 2016b; Rathbun et al., 2018) our map suggests that the poles of Io are comparably active to the equator. We compare our distribution of volcanic hotspot flux to the simulated surface heat flow distributions from de Kleer, Park, et al. (2019) and find that a uniform distribution of volcanic flux is more consistent with our flux map than any combination of these models. There may also be some magma scavenging controlling volcanic hotspot formation near the equator, as suggest by Hamilton et al. (2013) for the spacing between active hot spots. The global flux map also suggests some hemispheric dichotomy between the sub- and anti-Jovian hemispheres.

While these trends are interesting, it is worth mentioning that long-term ground-based imaging of Io has confirmed that its volcanism varies on many different timescales: by hours, days, months, years, and decades. While Juno can provide a complete global view of Io, it can only provide a snapshot of activity, and therefore cannot measure long-term trends and changes in activity. However, Veeder et al., 2012 suggests that outbursts contribute only 1%–2% of the total power emitted by Io and that a (slim) majority of all heat flow from Io can be accounted for by individual, active volcanic hotspots. Following this, the M-band brightness is likely representative of most volcanic activity that was active during the years of spacecraft- and ground-based observations.

**Figure 1.** The perspective shows the sub-Jovian, north polar view of Io in the left column and the anti-Jovian, south polar view of Io in the right column. The top-most row shows the coverage map achieved for JIRAM during this study. To reduce error from near-limb volcanic observations, we filter out all volcanoes whose emission angles are greater than 80°. The second row is a global map of volcanic flux measured in the M-band (4.78 μm) and given in mW/μm/m²/str. The extreme end members of deep mantle heating and asthenospheric/shallow mantle heating models are the 3rd and 4th rows. The asthenospheric heating model for a subsurface global magma ocean is the fifth row. These model end members are mapped with the total surface heat flow distribution in W/m² using modeling data from the 2019 Final Report for the Keck Institute for Space Studies (de Kleer, Park, et al., 2019). The relative distribution of heat flow is most important to compare to the volcanic flux distribution.

## 2. Methods

### 2.1. The JIRAM Instrument and Analysis

As the name suggests, the Jovian Infrared Auroral Mapper (JIRAM) instrument was not primarily intended to observe Io's volcanoes. JIRAM is also perpendicular to Juno's spin axis and, while Juno's unique polar orbit makes it incredibly useful for science exploration, it also makes the pointing of the camera complex which has been noted before (Adriani et al., 2017; Davies et al., 2024; Mura et al., 2020; Noschese et al., 2020; Zambon et al., 2023). Other studies have used this data set to study the locations of individual volcanoes, reporting both the location and the identity of each volcano (Davies et al., 2024; Zambon et al., 2023). The de-spinning mirror intended to counter Juno's relatively quick spin can only keep the image in the instrument field-of-view for a second. Nevertheless, the de-spun and fully processed data can measure the brightness of volcanoes and locate them with high enough accuracy to construct a complete global map.

JIRAM has two settings, a spectrometer and an infrared imager that observes in the L- and M-bands. We use only the M-band images (centered at 4.78 μm and integrated across a 480- nm bandwidth) on JIRAM which allows us to image Io's volcanoes, including the cooler, dimmer hotspots, with minimal competition from Jupitershine. In particular, M-band emission from a blackbody scales well with the total emission in the temperature range of silicate volcanism. This means that, assuming volcanoes emit as a blackbody, using M-band measurements are an excellent proxy for total emission. However, it should be acknowledged that volcanoes on Io are *not* blackbodies and their emission is not uniform. Large cooling lava flows often surround the hotter central hotspot, and there may be other regional variations in where the cooler, dimmer emission occurs versus where the hotter active radiation occurs (de Kleer et al., 2014; Rathbun et al., 2004; Rathbun & Spencer, 2010; Spencer et al., 1990; Tsang et al., 2014). With these caveats in mind, the M-band remains an excellent wavelength for observing Io's active volcanism and is routinely used for studies of Io's volcanism (Conrad et al., 2015; de Kleer and Rathbun, 2023). We have measured the individual locations and measured M-band power from each volcanic hotspot visible over 11 orbits (see Table S1 in Supporting Information S1) and used them to construct a globally complete map of volcanic flux (see Figure 1)– that is, the radiance emitted by volcanic hotspots (power per area, wavelength, and solid angle). Following the methods outlined by Mura et al. (2020), we first compile a "super-resolution" image from many snapshots taken during an orbit (see Figure S1 in Supporting Information S1). When arrays of images are separated by an interval of more than 30 s, these are treated as separate acquisition blocks; a unique super-resolution image is produced for each block in a single orbit. We use a fitting algorithm that locates the limb of Io and SPICE kernels to create an accurate latitude-longitude map across each image, creating a guide for volcano identification.

Unlike previous studies (e.g., Zambon et al., 2023) we calculate the brightness of each volcano, and we use only the brightness measured through individual images. Ultimately, the locations of individual volcanoes are used only for identification and to assign a central reference point. Since we use discrete volcanic hotspots to construct a global volcanic flux map, our map has a fundamental resolution that is dependent on the resolvable distance between individual volcanoes. In the orbit with the lowest resolution, the average distance between a volcano and its nearest neighbor is 18°, which gives us a baseline for our spatial resolution. Since our study focuses only on the relative distribution of volcanoes and the scale at which we evaluate variations is much higher than position and pointing error, we do not consider the uncertainty in the position of individual volcanic hotspots.

Our study includes a list of each detected hotspot in Supporting Information S1; however, our study does not use number density of individual volcanoes, but rather integrates brightness across these broad 18-degree counting circles. This negates any dependence on highly accurate pointing and latitude-longitude measurements and also makes direct comparison to the list of hotspots produced by Zambon et al. (2023) not meaningful.

First, we identify and establish the latitude and longitude of a volcano using the super-resolution image. To calculate the flux emitted from each volcano, we process and measure its brightness in the individual images that contain it during an orbit. However, many of the observations capture Io in partial daylight. Furthermore, the camera experiences "bleed," where particularly bright pixels can cause spillover to adjacent pixels (these artifacts appear as crosses around the brightest volcanoes. See Figure S1 in Supporting Information S1). We circumvent these issues by carefully hand-selecting each pixel belonging to each volcano in each image. This includes the spillover pixels associated with a particular volcano. If any volcano is observed in full or partial reflected sunlight, we also hand-select a representative background pixel and subtract out the flux contributed by reflected sunlight.

We used the combined pixel brightness for each volcano to calculate the emitted radiance of the individual volcano. The images are already calibrated with each pixel's intensity given in units of GW/μm/m$^2$/str. We compute the area of each pixel as the square of the slant distance between the spacecraft and the pixel location multiplied by the resolution of the camera (238 μm$^2$). The pixel's brightness, then, is the given intensity multiplied by the pixel area and divided by the cosine of the emission angle.

Since the JIRAM instrument can only change integration times between blocks of images, to capture the dimmest volcanoes on Io, the brightest volcanoes often saturate. At counts above 12,000 Digital Numbers (DN), the response of the camera is non-linear and calculating the radiance of these pixels systematically underestimates the response. Above 16,000 DN, the value is entirely saturated and an accurate brightness for that volcano can rarely be recovered from the data. To circumvent this, if any pixels in the volcano have a DN above 15,000 (to remain safely below that limit), we remove that image's measurement from the volcano's average. We preferentially choose to exclude measurements of volcanoes in images where any of their pixels have a value above 12,000 DN. However, for many of the brighter volcanoes, there are pixels that fall between the range of 12,000–15000 DN in every image during an orbit. Mura et al. ([2020](#)) suggests not to use values above the radiance that, for a given exposure time, corresponds to 12,000 DN. However, while the readings are non-linear, we can still recover brightness from them. Empirical meta-analysis of the JIRAM instrument finds that the response is roughly halved above 12,000 DN, and we therefore apply a doubling correction to any between 12,000 and 15,000 to recover an actual brightness. For each volcano observed, there is at least one observation that has only pixel values below 15,000; this allows us to keep full coverage across Io without losing any data to saturation (this is a statistical correction for this data set and should preferably not used for different sub-sets of JIRAM data). Unlike previous studies using this data set (i.e., Davies et al., [2024](#); Zambon et al., [2023](#)) our brightness measurements rely on individual frames and not super-resolution images, allowing us to remove saturated instances of volcanoes. Furthermore, we correct for saturation at a reliable threshold and in a well-understood regime, allowing for better constrained brightness measurements.

However, saturation issues impact recent orbits much less (or not at all) than earlier orbits (see in Supporting Information S1), a side effect of developing an Io hot spots research campaign that prefers finding new and faint hot spots over detecting the radiance of large and known ones. For this reason, the most recent orbits are the most reliable, and we recommend using data that does not rely on this correction for future studies.

## 2.2. Production of a Global Flux Map

To turn our discrete volcanic hotspot observations into a global map of volcanic hotspot flux, we use a technique commonly used in crater counting by averaging volcanic activity over a set area (Rathbun et al., [2018](#)). Since we use discrete volcanic hotspots to construct a global volcanic flux map, our map has a fundamental resolution that is dependent on the resolvable distance between individual volcanoes. For each point on our map, we add together the contributed volcanic brightness of each volcano within an 18-degree cone angle (the baseline resolution we use as described in the section above). We then divide that point by the area corresponding to an 18-degree spherical cap, given by

$$A = 2\pi r^2(1 - \cos\theta)$$

and, on Io, is equal to just over 1 million square kilometers. This gives us an average flux for that portion of the surface, which is then assigned to that single point of the map. This process is repeated for each point on the global flux map. The observed area of many of the orbits overlap (as shown in the Figure [1](#) above) and to account for repeated measurements of the same volcanic hotspots (and make some effort toward representing the time-averaged behavior), we create a separate coverage map of Io that counts how many orbits viewed Io within an 80-degree disk. We considered only the central 80 degrees of coverage instead of the full 90 degrees of disk coverage to help mitigate errors produced by near-limb observations of Io's volcanoes. Meta-analysis of our data established that the standard deviation in the brightness of each volcano was much higher image-to-image than the error introduced by emission angle uncertainty. However, as near-limb volcanic hotspots cannot necessarily be imaged in their entirety (and there is no way to guarantee that the entire volcano is captured), we still chose to exclude near-limb observations. All uncertainties reported are the standard deviation of brightness across each image in an orbit.

Using only the most accurate, face-on observations of the disk of Io, we still achieved global coverage of Io (including the elusive poles).

### 2.3. Spherical Harmonic Decomposition

To quantitatively assess the distribution of volcanic activity, we use spherical harmonic decomposition—that is, we fit our map to a map reproduced using spherical harmonic terms. Since we compare our data to the models in de Kleer, Park, et al. (2019), we use the minimum degree that creates a good match with all three models; in this case, both asthenosphere and deep mantle end members can be well-reproduced using terms up to degree $l = 4$, while the global magma ocean requires terms up to $l = 6$ for a close match. We generate the Legendre polynomials associated with each harmonic and use it to find the usual spherical harmonics; see the in Supporting Information S1. We are then left with an array for each node of spherical harmonics up to order $l = 6$. We vectorize these arrays and use them as independent variables in a multiple linear regression. The result is a vector of coefficients that describe what contribution from each harmonic creates the closest fit to our map.

Asynchronous rotation has been suggested on Io in previous studies (Abrahams et al., 2021; de Kleer & de Pater, 2016a; Hamilton et al., 2013; Park et al., 2019) which could create a longitudinal offset in the peak and lowest volcanic activity. To address this, we allow free rotation of the volcanic flux map to assess whether a longitudinal offset, and how large of an offset, could provide the best fit. To do this, we perform multiple linear regressions across a full rotation of Io and choose the fit with the lowest residuals.

## 3. Results

### 3.1. Notable Volcanoes

Our study produced a globally complete map of the flux produced by volcanic hotspots; viewing this flux on both a linear and a logarithmic scale better illustrates individual volcanic behavior and global heat flow variations, particularly the lowest-flux regions (see Figure 2). This map includes several interesting bright volcanic features. While we do not attempt to identify all the volcanoes we measure, we were able to determine identities for these particularly bright features.

The bright circle centered near 300 W, 20 N is Loki, the brightest and largest persistent volcano on Io. It is surrounded by an area of noticeably low volcanic flux especially on its equatorial side. Near 60 N, 80 W are four bright volcanoes: Asis, Zal, Tonatiuh, and an unnamed extended volcanic source at 51 N, 87 W. These four clustered hotspots are visible and highly active (forming a distinctive "L" shaped feature recognizable from orbit to orbit; see Figure S1 in Supporting Information S1) throughout each fly-by that imaged that region between 2018 and 2020. By the last encounter in this data set in 2020, the feature had dimmed from a combined radiance of 10.2 GW/μm/str (Orbit 11) to 7.4 GW/μm/str (Orbit 27), but it continued to be visible at comparable levels through the entire data set. Asis was first observed in 2001 (Lopes et al., 2004) and Zal in 1996 (Lopes- Gautier et al., 1999); Tonatiuh's first observation was in 2010 (de Pater et al., 2014) (although it may have been detected earlier, during the Galileo mission). The JIRAM detections of these volcanoes extend their observation history to over 19 years of activity. Furthermore, since they are all visible in the 4-year observation period, all four meet the 1-year consistent activity requirement from de Kleer et al. (2016b)– all four are persistent hotspots (it's worth noting that the second requirement to be a persistent hotspot regards a maximum L'-band brightness of 15 GW/μm/str. Since their intensities were measured only in M-band, we simply state that none of these volcanoes have been observed outbursting). These high-latitude volcanoes should therefore be considered when calculating the statistical relative spatial distribution of persistent hotspots compared to sudden eruptions.

Nearby to the L-quartet is Tvashtar, a volcano which has exhibited several outbursts without being observed at a low level (de Kleer & Rathbun, 2023). During the Galileo period, Tvashtar was observed as an outburst by both Galileo and ground-based telescopes (Howell et al., 2001; Lopes et al., 2004; McEwen et al., 2000; Milazzo et al., 2005). It was again observed in outburst in 2006/2007 by both ground-based telescopes and the New

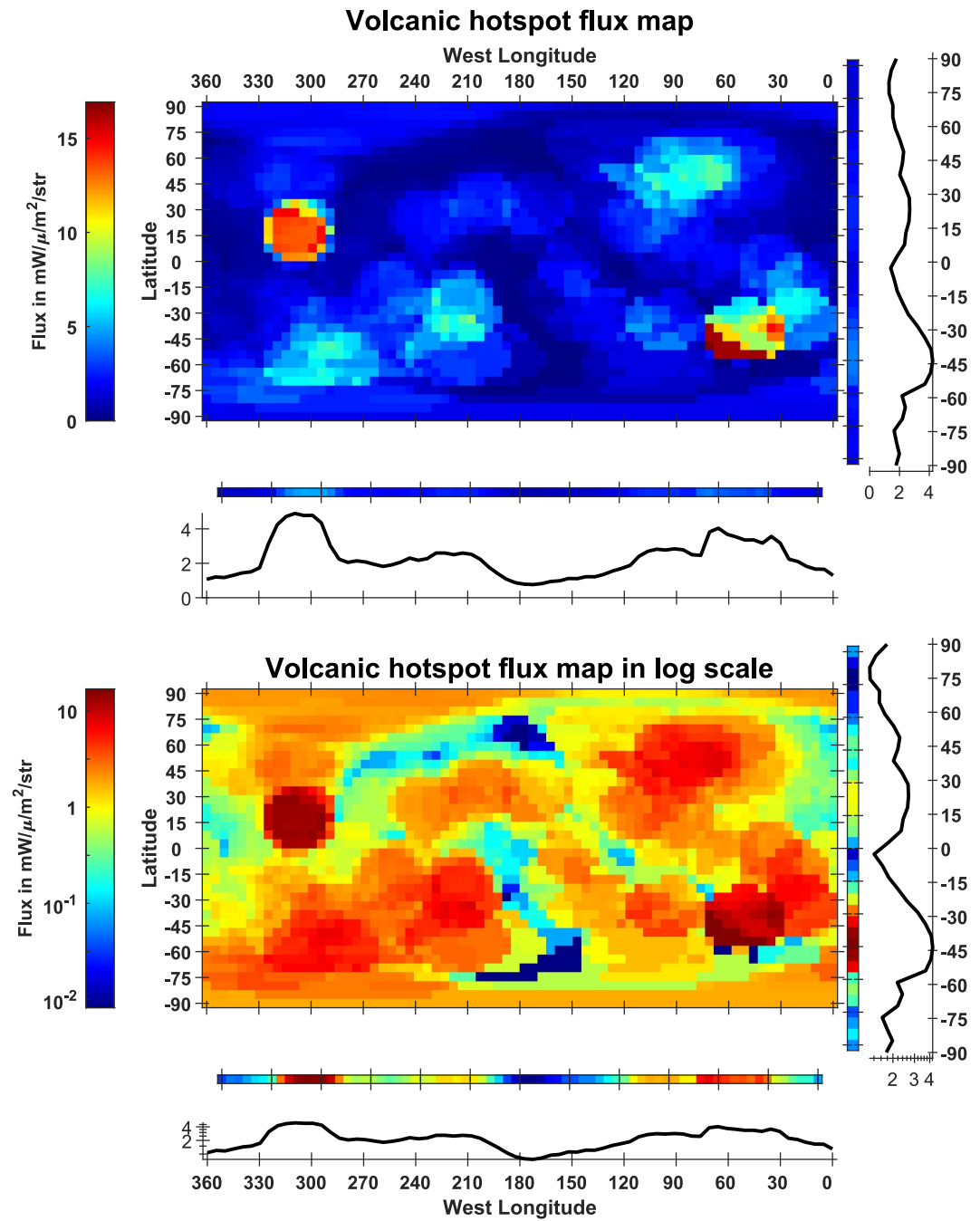

**Figure 2.** These are maps of the global volcanic flux on Io in an equirectangular projection, showing the averaged volcanic flux in milliwatts per square meter per steradian (the most common units for volcanic flux on Io). The top is on a linear scale while the bottom is on a logarithmic color scale. The colored bars and the line plots beside each map show the average flux projected horizontally (to the right of each map) and the average flux projected vertically (below each map) to show trends in flux by latitude and longitude.

Horizons spacecraft (Laver et al., 2007; Spencer et al., 2007). JIRAM imaged Tvashtar in Orbits 17, 18, and 26, which was also noted by Davies et al., 2024. By Orbit 26 (which occurred in April of 2020) however, it was quite dim at only 0.83 GW/μm/str. This was the first time Tvashtar was observed at a low level of activity, perhaps because, due to its high latitude, only very bright eruptions would be observable from the previous observations which were equatorially-biased.

The third system of bright volcanoes on Io is in the southern leading hemisphere of Io, centered near 30 W, 40 S. Three bright volcanoes form this feature: Kanehekili, Laki-oi, and Uta, previously observed in ground-based campaigns (de Kleer, Nimmo, et al., 2019; Rathbun & Spencer, 2010). Kanehekili was observed with a peak M-band brightness of 74 ± 12.0 GW/μm/str in 2010 (de Pater et al., 2014).

As part of this study, we considered whether any outbursts were observed by JIRAM. However, the canonical definition of an outburst proposed by Sinton et al. (1982), revised by Stansberry et al. (1997) and later by de Kleer and de Pater (2016a), and most recently formalized by Tate et al. (2023) uses the L' brightness and not the M-band brightness. Mini-outburst levels are similarly based on L' measurements. However, none of the volcanoes we observed reached what we would consider mini-outburst or outburst level in flux, the highest single point-source power being lower by an order of magnitude. However, we found that the brightest hotspots (which, based on the behavior of our data, we chose to assign to volcanoes with a power above 10 GW/μm/str) are concentrated about the areas of highest volcanic activity: Loki, the northern leading quartet of volcanoes, and the southern leading collection of volcanoes all have multiple observations at above 10 GW/μm/str.

### 3.2. Global Flux Map

The areas of highest volcanic flux appear to be surrounded by areas of low flux (particularly notable around Loki, but present across the equator and mid-latitudes). This could suggest some magma scarcity (Hamilton et al., 2013); in an area of high volcanic activity, the shared magma reservoirs could be depleted and decrease the possibility of volcanic activity nearby (or other processes, like the introduction of extensional stress in the lithosphere near areas of high volcanic activity that could inhibit the formation of dikes, and therefore inhibit the development of other volcanoes; e.g. in Hamilton et al. (2013), McGovern et al. (2016)).

The areas of lowest volcanic flux seem to be concentrated at the anti-Jovian hemisphere or offset by tens of degrees west of the anti-Jovian point in the trailing hemisphere. There does appear to be a hemispheric dichotomy in behavior: the leading hemisphere is more uniformly bright and has multiple bright volcanic features concentrated at Tvashtar and the L-quartet, whereas the trailing hemisphere has a single bright volcanic source (Loki) surrounded by very dim and scattered hotspots.

In order to compare our results to those of Davies et al. (2024) (who analyzed the same data set using different methodology), we determined the total output at the poles to see if there is a north-south hemispheric dichotomy. Above 60 N and below 60 S, we find a discrepancy in activity—the north is 65% brighter than the south pole; a combined radiance of 7.2 TW versus 1.6 TW (a discrepancy is also measured in Davies et al., 2024). However, unlike Davies et al., we do not consider this difference in output significant. We note that any separation between equatorial and polar regions is somewhat arbitrary (and results change if the separation latitude of 60° is changed). The south pole was only observed twice (see Figure 1 above) while the north pole was observed nine times– since many of Io's volcanoes fluctuate on short timescales, the few snapshots taken at the south pole are contingent on the conditions at the time of the fly-by (in this case, Orbits 10 and 25, which occurred in late 2017 and early 2020 respectively). Furthermore, since orbit 10 was an early orbit and therefore suffered more saturation than later orbits (see Supporting Information S1), there is a greater uncertainty in the brightness of the south pole. Therefore, the discrepancy in total brightness between the north and south poles may not be representative of general trends.

### 3.3. Comparisons to Tidal Heating Models

Once we produce a global map of volcanic flux, we compare our map to the surface heat flow maps created by modeling the most extreme cases of the tidal heating emplacement summarized by de Kleer, Park, et al. (2019). These models suggest that the depth at which tidal heat is emplaced would create measurable differences in the spatial distribution of surface heat flow. The models explore heat emplacement from shallow (asthenospheric) to deep (deep mantle) and additionally account for the longitudinal offset that a lithosphere decoupled from the mantle via a globally extensive magma ocean (global magma ocean) would create. The deep mantle and asthenospheric heating models were first presented in Segatz et al. (1988), while the global magma ocean model presented in de Kleer, Park, et al. (2019) was made by combining the models of Tyler et al. (2015) and Matsuyama et al. (2018). These case end members are shown in Figure 1. Juno's polar orbit is of critical importance for an accurate comparison between the modeled data and volcanic flux measurements. As can be seen in Figure 1, the relative surface heat flow of the model end members differs most strongly in the relative flow through the poles versus the equator. This makes accurate flux measurements of the poles necessary for any robust discrimination between model regimes.

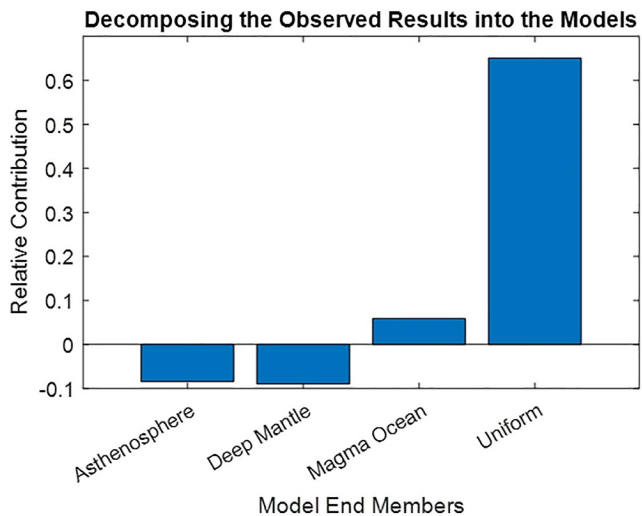

**Figure 3.** This plot shows the results of reconstructing the observed global volcanic flux map using linear combinations of the extreme end members of the models. The coefficients for the best fit to the observed volcanic flux are above. A uniform map is most consistent with the observed volcanic flux. Of the model end members, the global magma ocean heating regime has the highest correlation with the observed flux, while the deep mantle and asthenosphere are both anti-correlated with the observed flux.

### 3.3.1. Results of Spherical Harmonic Decomposition

Using the methods described in Section 2.3, we decompose both our accrued data and the tidal heating models into their composite spherical harmonics. We then compare the degree to which the coefficients of each harmonic agree between the reconstruction of our data and the reconstruction of the models. We found that the observed volcanic hotspot flux could not be neatly described by any of the model end members, even when allowing for free rotation. Each individual tidal heating regime end member did not fit the observed flux well—see Figure S2 in Supporting Information S1. Of particular note is the $l = 2$, $m = 0$ harmonic, where there are strong contributions from all three models, but very little contribution to the observed flux map. Similarly, $l = 3$, $m = -2$ strongly contributes to the global flux map while none of the models have a corresponding component. Importantly, there is not strong agreement between the terms that make up the models and the map, including in the lower-order $l = 0$–2 harmonics. This leads us to conclude that none of the models strongly agree with the observed flux.

Our spherical harmonic fitting routine allows for free rotation of Io; that is, if there exists a better fit when Io is rotated latitudinally or longitudinally, the routine will account for that. This means that the disagreement with models is not the result of lithosphere-mantle decoupling or an offset in the patterns, but rather a disagreement with any of the expected distributions.

### 3.3.2. Decomposition Into Model End Members

While the spherical harmonic decomposition suggests that there is not good agreement with the model, we assessed which model best fits the data and to what extent they agree. Quantitatively comparing our global volcanic flux map to these end members, and crucially the degree to which the real global data matches these predicted outcomes, allows us to test which surface heat flow regime (or which linear combination of heat flow regimes) best describes the observed volcanic flux.

To do this, we reconstructed our observed global flux map using the three model end members and a uniform distribution model (i.e., a map with equal flux everywhere) as a parameter space. This is similar to the method described in Section 2.3, where we vectorize the maps and use them as variables in a multiple linear regression. While these model end members are not orthogonal and therefore cannot form a true basis, it gave us a sense of how a combination of the tidal heating depths and magma ocean layer might contribute to the observed volcanic flux. We found that the observed flux on Io is more consistent with a uniform distribution than any of the model end members which reiterates that the model does not fully capture the volcanic distribution on Io (see Figure 3 below). However, of the model end members, the observed volcanic flux has the highest correlation with the global magma ocean heating model, but only because it has an anti-correlation with both the deep mantle and asthenosphere model. All these correlations are incredibly small compared to the correlation with a uniform distribution. Since the routine allows for free rotation and therefore accounts for longitudinal offsets, we note that the anti-correlation of the asthenosphere and deep mantle models is not caused by a longitudinal offset.

The global volcanic flux map has lower activity at the poles, but the areas of the most anomalously low flux occur at the equator and at the sub- and anti-Jovian points. We believe that reconstructing the observed map requires replicating the areas of low heat flow at the equator (which leads to an anti-correlation in the global magma ocean end member, whose peak heat flow occurs at the equator). The anti-correlation with the deep mantle model is likely also due to the lower volcanic flux at the poles. The areas of highest volcanic flux are at the mid-latitudes, which is qualitatively consistent with the asthenospheric heating regime. However, we again stress that none of the models provide a good fit to the observed distribution.

## 4. Discussion

JIRAM and Juno have offered an unprecedented view of Io and have completed our global image of Io's volcanism. We compiled a database of saturation-corrected individually resolved volcanic hotspots imaged over multiple orbits and used it to construct a global map of the flux produced by volcanic hotspots, including the under-observed polar regions. This has offered critical new insight toward the behavior of volcanism and the spatial distribution of volcanoes on Io. We have identified several key areas of volcanic activity that are of particular interest. The first is an L-shaped quartet of mid- and high-latitude volcanoes, all of which were visible for the entirety of this study. Two of the four, Zal and Tonatiuh, were classified as persistent by de Kleer and de Pater ([2016a](#)). Another was Asis Patera, where activity was also detected by Lopes et al. ([2004](#)), de Kleer and de Pater ([2016a](#)) (originally referred to as Aluna). However, at the time, it did not meet de Kleer et al.'s definition of persistent. The fourth spot was observed by Mura et al., [2020](#) and is located north of Zal and may be part of the same complex, but given the high spatial resolution of the JIRAM data, appears to be a separate hotspot. Given that all four volcanoes, including Asis, were consistently detected through this 4-year observation period, we suggest that the quartet each meet the definition for persistent volcanism. Tvashtar, which is a high-latitude volcano at 62 N, was observed to be active in at least three orbits during this study (the latest of which occurred in 2020). Tvashtar has been imaged since 1999 (Milazzo et al., [2005](#)) both in quiescent periods and during outburst (Lopes et al., [2004](#); McEwen et al., [2000](#); Spencer et al., [2007](#)).

McEwen et al. ([2000](#), [2004](#)) suggested that while longer-lived eruptions occur at low latitudes with shorter duration, higher volume eruptions occur at high latitudes. de Kleer et al., [2016a](#), [2016b](#) also found that persistent hotspots occur primarily within 30 degrees of the equator. Further observations of Io's high latitude regions, and these five volcanoes, will give insight into the latitudinal distribution of volcano types and address the different definitions that have been used for "persistent" or "long-lived" volcanoes.

Areas of highest volcanic activity are loosely concentrated at the mid-latitudes. The areas of Io with M-band powers greater than 10 GW/µm/str during this study were concentrated in three regions: Loki, the L-quartet (Asis, Tonatiuh, Zal, and a fourth unnamed source), and the southern leading hemisphere trio (Kanehekili, Uta, and Laki-Oi). Spherical harmonic analysis does not show a strong agreement between any of the modeled heat flow regimes and the observed volcanic activity. This can be explained by a number of physical processes and could suggest that the distribution of volcanic flux is not comparable to the surface heat flow (Stevenson & McNamara, [1988](#)).

Previous studies have measured higher volcanic activity at the equator (Rathbun et al., [2018](#)), which they interpret to be consistent with asthenospheric heating (de Kleer et al., [2016b](#)). Davies et al., [2024](#), which uses the same data set, finds that the total brightness above 60 N and below 60 S is lower than between 60 N and 60 S and interpret this as being consistent with asthenospheric heating. Our study corroborates lower activity at the poles but disagrees with the interpretation that this is consistent with asthenospheric heating.

Our analysis finds that the observed M-band flux is anti-correlated with the asthenospheric heating model and has only very weak agreement with the global magma ocean heating model. Using spherical decomposition, we find that the distribution of flux is much more uniform than in-line with any of the models. It is anti-correlated with the asthenospheric heating model, and we therefore reject this conclusion. Areas of low volcanic activity are concentrated about 180 W and near the equator, which is consistent with none of the end members but is anti-correlated with both the asthenospheric and deep mantle models. Our study finds that an offset is not sufficient to account for this anti-correlation and that other models of spatial control may be necessary to explain the observed volcanic activity.

One such example of spatial control is Loki, which is the brightest hotspot on Io but is noticeably much brighter than the entire trailing hemisphere. The areas surrounding Loki to the south are particularly dim in comparison to the average volcanic flux of Io, which suggests there may be some magma scavenging behavior (Hamilton et al., [2013](#)). Shared magmatic sources could be drained by Loki's high activity. Loki's volcanism may also cause extensional stress in the lithosphere that prevents the formation of other dikes nearby. This could also account in some part for an apparent hemispheric dichotomy between the leading and trailing hemispheres. Aside from Loki, the trailing hemisphere has fewer high-activity clusters; its activity seems to be concentrated on Loki, opposed to the leading hemisphere which seems more uniformly active.

Other studies have used the JIRAM data set to analyze the distribution of volcanic hotspots (Davies et al., 2024; Zambon et al., 2023). Unlike those studies, we do not use super-resolution images to measure the brightness of hotspots, and instead measure and average the brightness over its composite frames. This allows us to be much more discerning in which images we use in an averaged measurement—preferentially removing any instance of saturated data or, when unsaturated data is not available for a specific hotspot, correcting those measurements within a well-understood regime (described in detail in Section 2.1). Both Zambon et al., 2023; Davies et al., 2024 provide a list of identified hotspots. We corroborate that there are at least several new observations of high-latitude volcanoes as is described in both papers but cannot compare to the distribution of individual hotspots discussed above as we do not have the spatial resolution to identify each hotspot we observe, outside of a few particularly bright volcanoes discussed above. A list of our measured hotspot locations and brightnesses is in the Supporting Information S1. Furthermore, unlike Davies et al., 2024, we do not interpret our flux distribution to be consistent with the asthenospheric heating regime. While we agree that the flux produced above 60 N and below 60 S is less than the flux produced between 60 S and 60 N, we find that decomposing our map either into its composite spherical harmonics or into the model end members both show the same result: no model (or combination of models) matches the observed flux well.

Juno and JIRAM have allowed us to compile a global view of Io and analyze spatial trends in volcanic activity. While our study includes global coverage of Io, the snapshots provided of Io are highly dependent on the timing of fly-bys. Unlike ground-based observing campaigns, this study has compiled data over a relatively short timescale (4 years) with relatively few observations (11). If volcanic activity has significantly changed in spatial distribution as some recent studies suggest (e.g., Tate et al., 2023), this study is representative only of Io's most recent behavior. As Juno continues its mission, we will continue our global study of Io's volcanism while providing a more time-averaged view of its behavior. Further study may reveal a more coherent narrative for the overall trends in high- and low-volcanic activity and allow for the more in-depth analysis of individual volcanic hotspots.

## Data Availability Statement

The data used in this study comes from Juno and the Jovian Infrared Auroral Mapper (JIRAM) and is publicly available on the Planetary Database System in PDS4 format (Adriani et al., 2019). The SPICE bundles used to calibrate the viewing geometry is also publicly available in the Planetary Database System via the Navigation and Ancillary Information Facility (NAIF). Adriani et al. (2019), Juno JIRAM Bundle, [Dataset] PDS Atmospheres (ATM) Node, https://doi.org/10.17189/1518967.

**Acknowledgments**

This work was funded by the New Frontiers Data Analysis Program under Grant 80NSSC19K1093 and by the New York Space Grant. Many thanks to the JIRAM team for making this work possible. Thank you to Isamu Matsuyama, who provided us the maps of surface heat flow that were presented in the Park and K. de Kleer et al., 2019 KISS Conference report.